\documentclass[pacs]{revtex4}
\usepackage{amsmath}
\usepackage{amssymb}
\usepackage{graphicx}
\begin{document}

\title{Classical color fields as a dark matter candidate}
\author{Vladimir Dzhunushaliev
\footnote{Senior Associate of the Abdus Salam ICTP}} 
\email{dzhun@krsu.edu.kg} 
\affiliation{Dept. Phys. and Microel. 
Engineer., Kyrgyz-Russian Slavic University, Bishkek, Kievskaya Str. 
44, 720021, Kyrgyz Republic}


\begin{abstract}
The model of Dark Matter is proposed where the Dark Matter is a classical color field. The color fields are invisible as they may interact with colored elementary particles like 't Hooft - Polyakov monopole only. The comparison with the Universal Rotation Curve is carried out. 
\end{abstract}

\maketitle

\section{Introduction}

In astrophysics and cosmology, Dark Matter (DM) is matter of unknown composition that does not emit or reflect enough electromagnetic radiation to be observed directly, but whose presence can be inferred from gravitational effects on visible matter. According to present observations of structure larger than galaxy-sized as well as Big Bang cosmology, DM accounts for the vast majority of mass in the observable universe. Among the observed phenomena consistent with DM  observations are the rotational speeds of galaxies and orbital velocities of galaxies in clusters, gravitational lensing of background objects by galaxy clusters such as the Bullet cluster, and the temperature distribution of hot gas in galaxies and clusters of galaxies. DM  also plays a central role in structure formation and galaxy evolution, and has measurable effects on the anisotropy of the cosmic microwave background. All these lines of evidence suggest that galaxies, clusters of galaxies, and the universe as a whole contain far more matter than that which interacts with electromagnetic radiation: the remainder is called the ``dark matter component''.
\par
The first to provide evidence and infer the existence of a phenomenon that has come to be called ``dark matter'' was F. Zwicky \cite{Zwicky}. He applied the virial theorem to the Coma cluster of galaxies and obtained evidence of unseen mass. 
\par 
According to results published in \cite{Clowe:2006eq}, dark matter has been observed separate from ordinary matter through measurements of the Bullet Cluster, actually two nearby clusters of galaxies that collided about 150 million years ago. Researchers analyzed the effects of gravitational lensing to determine total mass distribution in the pair and compared that to X-ray maps of hot gases, thought to constitute the large majority of ordinary matter in the clusters. The hot gases interacted during the collision and remain closer to the center. The individual galaxies and DM did not interact and are farther from the center. Recently the Bullet cluster data offered as evidence of dark matter 
\par
The numerous astrophysical observations, e.g., Doppler measurements of rotation velocities in disk galaxies, have established the failure of the classical Newtonian theory, if only visible matter is taken into account \cite{Combes95}. Historically, theoretical concepts addressing this problem can be subdivided in two categories. The first category comprises the DM theories \cite{BiTr94}, whereas the second group assumes that Newton's gravitational law requires modification \cite{Milgrom:1983ca}. 
\par 
DM theories are based on the hypothesis that there exist significant amounts of invisible (non-baryonic) matter in the universe, interacting with ordinary visible matter only via gravity. Since empirically very successful, DM has become a widely accepted cornerstone of the contemporary cosmological standard model \citep{Sa99}. Nevertheless, it must also be emphasized that until now DM has been detected only indirectly by means of its gravitational
effects on the visible matter or the light.  
\par 
In order to explain the existence of region where the velocity of stars is 
$\approx const$ it is necessary (in the DM framework) to have lengthly enough a region with DM. For the classical SU(3) gauge theory it is the ordinary situation: the point is that the classical Yang-Mills equations ordinary give us the solutions with an infinite mass \cite{Obukhov:1996ry}, i.e. the mass density is those that $\rho \geq r^{-2}$. By indirection this problem is connected with the confinement problem in quantum chromodynamics which claims that the field distribution between quark and antiquark is a flux tube with a finite linear energy density. Such flux tube can not be obtained in the framework of the classical Yang-Mills theory. It is possible in the framework of the quantum Yang-Mills theory only. 
\par
The classical gauge theories are a nonlinear geberalization of Maxwell electrodynamics. The non-Abelian gauge theories were invented by Yang and Mills in 50$^{th}$ of the preceding century. For most of this period it was not known whether any of the interactions obseved in nature can be described by a non-Abelian gauge theory. Nevertheless, the elegance of these theories attracted interest. The Weinberg-Salam model ($SU(2) \times U(1)$ gauge theory) and quantum chromodynamics ($SU(3)$ gauge theory) are the two existing Yang-Mills theories of real phenomenological importance. These theories can be formulated in terms of Feynman path integrals, i.e. functional integrals over all classical field configurations weighted by a factor $\exp(-\mathrm{action})$. If one knew everything about classical field configurations, then in principle all questions concerning the quantum theory could be answered. Partial information about classical fields might yield, at least, some insight into the quantum theory. For review of classical solutions of Yang-Mills theories see Ref. \cite{Actor:1979in}.
\par
In this paper we use the solutions of the classical SU(3) gauge theory for the explanation of the rotational velocity of stars outside the core of the galaxy. 

\section{Initial equations for the SU(3) gauge field}

We consider the classical SU(3) Yang-Mills gauge field $A^a_\mu$. The field equations are \begin{equation}
	D_\nu F^{a\mu \nu} = 0 
\label{1-10}
\end{equation}
where $F^a_{\mu \nu} = \partial_\mu A^a_\nu - \partial_\nu A^a_\mu + g f^{abc} A^b_\mu A^c_\nu$ is the field strength tensor; $f^{abc}$ are the SU(3) structural constants; 
$a,b,c = 1, 2, \cdots , 8$ are color indices; $g$ is the coupling constant. We use the following ansatz \cite{corrigan} 
\begin{eqnarray}
	A_0^2 &=& - 2 \frac{z}{gr^2} \chi(r), \quad
	A_0^5 = 2 \frac{y}{gr^2} \chi(r), \quad
	A_0^7 = - 2 \frac{x}{gr^2} \chi(r), 
\label{1-30}\\
	A^2_i &=& 2 \frac{\epsilon_{3ij} x^j}{gr^2} \left[ h(r) + 1 \right] ,
\label{1-40}\\
	A^5_i &=& -2 \frac{\epsilon_{2ij} x^j}{gr^2} \left[ h(r) + 1 \right] ,
\label{1-50}\\
	A^7_i &=& 2 \frac{\epsilon_{1ij} x^j}{gr^2} \left[ h(r) + 1 \right] 
\label{1-60}
\end{eqnarray}
for the $SU(2) \in SU(3)$ components of the gauge field and 
\begin{eqnarray}
	\left( A_0 \right)_{\alpha , \beta} &=& 2 \left( 
		\frac{x^\alpha x^\beta}{r^2} - \frac{1}{3} \delta^{\alpha \beta}
	\right) \frac{w(r)}{gr} ,
\label{1-80}\\
	\left( A_i \right)_{\alpha \beta} &=& 2 \left(
		\epsilon_{is \alpha} x^\beta + \epsilon_{is \beta} x^\alpha
	\right) \frac{x^s}{gr^3} v(r) ,
\label{1-90}
\end{eqnarray}
for the coset components; $i=1,2,3$ are space indices; $\epsilon_{ijk}$ is the absolutely antisymmetric Levi-Civita tensor; the functions $\chi(r), h(r), w(r), v(r)$ are unknown functions. The coset components $\left( A_\mu \right)_{\alpha \beta}$ in the matrix form are written as 
\begin{equation}
	\left( A_\mu \right)_{\alpha \beta} = 
	\sum \limits_{a=1,3,4,6,8} A_\mu^a \left( T^a \right)_{\alpha , \beta} 
\label{1-110}
\end{equation}
where $T^a = \frac{\lambda^a}{2}$ are the SU(3) generators, $\lambda^a$ are the Gell-Mann matrices. The corresponding equations are 
\begin{eqnarray}
	x^2 w'' &=& 6w \left( h^2 + v^2 \right) - 12 h v \chi,
\label{1-120}\\
	x^2 \chi'' &=& 2 \chi \left( h^2 + v^2 \right) - 4 h v w,
\label{1-130}\\
	x^2 v'' &=& v^3 - v + v \left( 7 h^2 - w^2 - \chi^2 \right) + 2h w \chi, 
\label{1-140}\\
	x^2 h'' &=& h^3 - h + h \left( 7 v^2 - w^2 - \chi^2 \right) + 2 v w \chi 
\label{1-150}
\end{eqnarray}
here the dimensionless radius $x = r/r_0$ is introduced, $r_0$ is a constant. 

\section{Numerical investigation}

\begin{figure}[h]
\begin{minipage}[t]{.45\linewidth}
  \begin{center}
  \fbox{
  \includegraphics[height=5cm,width=7cm]{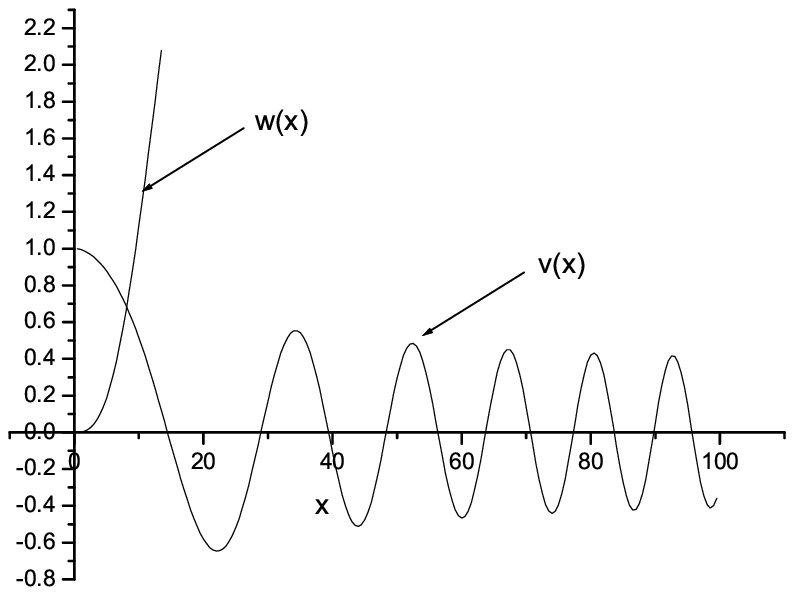}}
  \caption{The profile of functions $w(x), v(x)$, $v_2=10^{-2}$, $w_3=10^{-2}$.}
  \label{fg1}
  \end{center}
\end{minipage}\hfill
\begin{minipage}[t]{.45\linewidth}
  \begin{center}
  \fbox{
  \includegraphics[height=5cm,width=7cm]{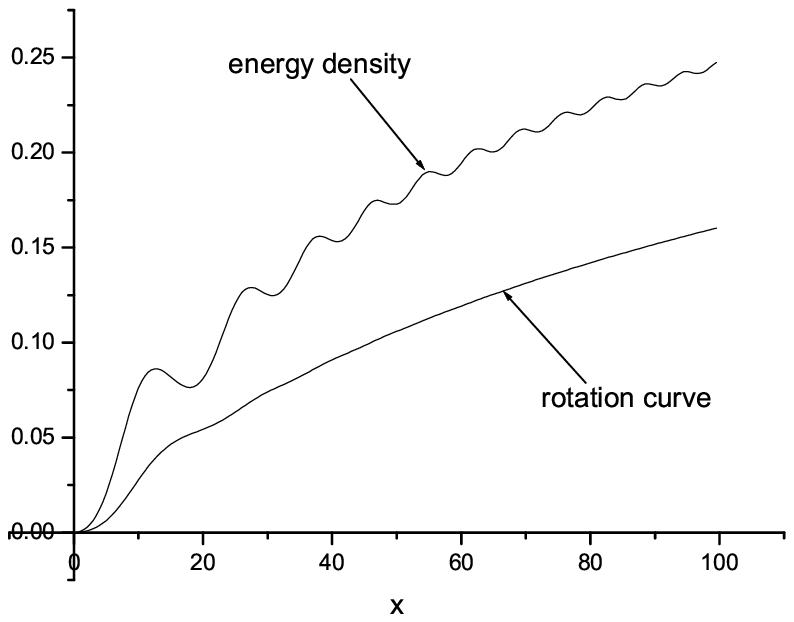}}
  \caption{The profile of the dimensionless energy density $\varepsilon(x)$ and the 							dimensionless rotation curve $\frac{c {g'}^2 r_0^2}{G \hbar} V^2(x)$.}
  \label{fg2}
  \end{center}
\end{minipage}\hfill 
\end{figure}

In this section we present the typical numerical solution of Eq's \eqref{2-10} \eqref{2-15}. 
We will investigate the case $\chi = h = 0$
\begin{eqnarray}
	x^2 w'' &=& 6w v^2 ,
\label{2-10}\\
	x^2 v'' &=& v^3 - v - v w^2 . 
\label{2-15}
\end{eqnarray}
For the numerical investigation we have to start from the point $x = \delta \ll 1$. Here we have the following approximate solution 
\begin{equation}
	v \approx 1 + v_2 \frac{x^2}{2}, \quad 
	w \approx w_3 \frac{x^3}{2}, \quad x \ll 1
\label{2-20}
\end{equation}
where $v_2, w_3$ are arbitrary constants. The typical behavior of functions $v(x)$ and $w(x)$ is presented in Fig.~\ref{fg1}. 
\par
The mass density $\rho(x)$ is 
\begin{equation}
	\rho = \frac{1}{2 c^2} \left(
	- F^a_{0i} F^{a0i} + \frac{1}{4} F^a_{ij} F^{aij} 
	\right) = 
	\frac{1}{g^2 c^2 r_0^4} \left[ 
	4 \frac{{v'}^2}{x^2} + 
	\frac{2}{3} \frac{\left( x w' - w \right)^2}{x^2} + 
	2 \frac{\left( v^2 - 1 \right)^2}{x^4} + 
	4 \frac{ v^2 w^2}{x^4} 
	\right] = 
	\frac{1}{g^2 c^2 r_0^4} \varepsilon(x)
\label{2-60}
\end{equation}
where $c$ is the speed of light. The profile of the dimensionless energy density  $\varepsilon(x)$ in Fig.~\ref{fg2} is presented.
\par
The rotation curve is defined as 
\begin{equation}
	V^2 = G \frac{m(r)}{r} = 
	\frac{4 \pi G}{c^2} \frac{1}{r}
	\int \limits^r_0 r^2 \rho(r) dr = 
	\frac{G \hbar}{c} \frac{1}{{g'}^2 r_0^2} \frac{1}{x} 
	\int \limits_0^x x^2 \varepsilon(x) dx = 
	\frac{G \hbar}{c} \frac{1}{{g'}^2 r_0^2} \frac{m(x)}{x}
\label{2-70}
\end{equation}
where $m(r)$ is the mass of the color fields $A^a_\mu$ under sphere with the radius 
$r = x r_0$, $m(x)$ is the dimensionless mass, ${g'}^2 = g^2 c \hbar$ is the dimensionless coupling constant, $G$ is the Newton gravitational constant. 
\par 
The asymptotical behavior of the solution $x \gg 1$ is 
\begin{eqnarray}
	v(x) &\approx& A \sin \left( x^\alpha + \phi_0 \right),
\label{2-80}\\
	w(x) &\approx& \pm \left[
		\alpha x^\alpha + \frac{\alpha - 1}{4} 
		\frac{\cos \left( 2 x^\alpha  + 2 \phi_0 \right)}{x^\alpha}
	\right] ,
\label{2-90}\\
	3 A^2 &=& \alpha (\alpha - 1)
\label{2-100}
\end{eqnarray}
with $\alpha > 1$.

\section{The comparison with a Universal Rotation Curve of spiral galaxies}

Unfortunately we have not any analytical solution and therefore it is very difficult to carry out the numerical investigation as the coefficient $\frac{G \hbar}{c}$ in Eq. \eqref{2-70} is very small and asymptotically we have strongly oscillating function $v(x)$ in Eq. \eqref{2-80} (if $r_0$ is very small). Therefore in this section we will investigate the rotation curve of gauge DM close to the center and far away from the center. 
\par
\begin{figure}[h]
\begin{minipage}[t]{.45\linewidth}
  \begin{center}
  \fbox{
  \includegraphics[height=5cm,width=7cm]{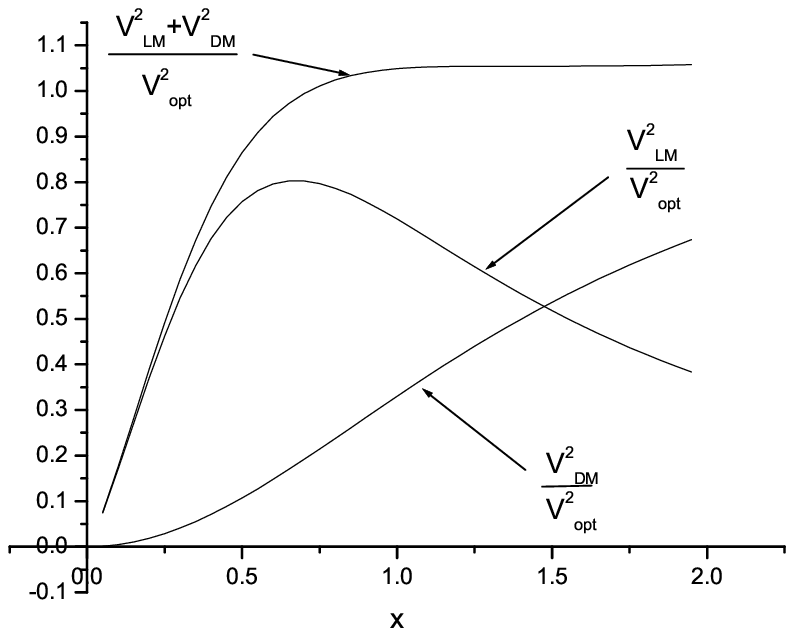}}
  \caption{The profiles of dimensionless rotation curves for the light and dark matter \cite{Persic:1995ru}.}
  \label{fg3}
  \end{center}
\end{minipage}\hfill
\begin{minipage}[t]{.45\linewidth}
  \begin{center}
  \fbox{
  \includegraphics[height=5cm,width=7cm]{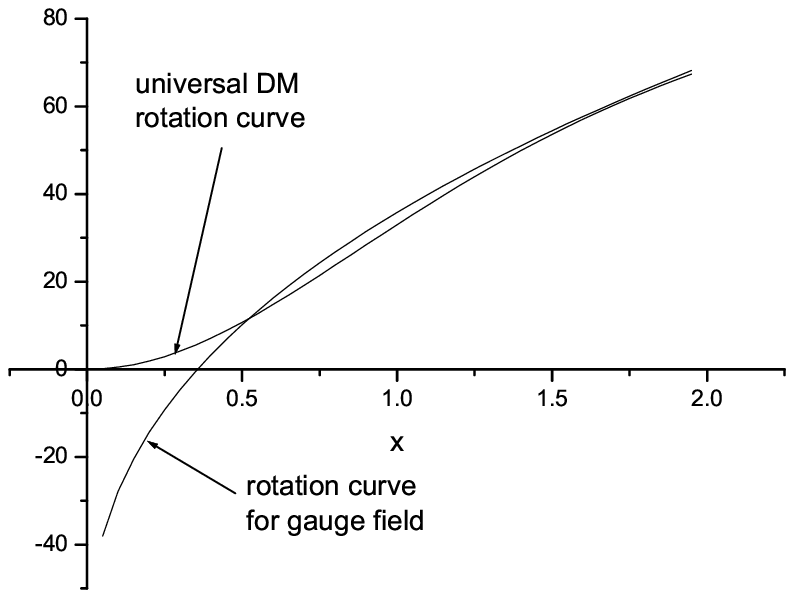}}
	  \caption{The comparison of DM rotation curve \eqref{3-30} with the rotation curve 							\eqref{3-80} for the SU(3) classical color field \eqref{1-30}-\eqref{1-90}. 
	  $\alpha = 1.2, g' = 0.3, R_{opt} = 20 KPs = 6*10^{17} m, V_{opt} = 100 Km/s = 10^5 m/s$,
	  $V_0 = 70 Km/s = 7*10^{4} m/s$.}
  \label{fg4}
  \end{center}
\end{minipage}\hfill 
\end{figure}

 
In Ref.~\cite{Persic:1995ru} a Universal Rotation Curve of spiral galaxies is offered which describes any rotation curve at any radius with a very small cosmic variance
\begin{equation}
	V_{URC} \left( \frac{r}{R_{opt}} \right) = 
	V(R_{opt}) \left[ 
		\left( 0.72 + 0.44 \log \frac{L}{L_*} \right) 
		\frac{1.97 x^{1.22}}{ \left( x^2 + 0.78^2 \right)^{1.43}} + 
		1.6\, e^{-0.4(L/L_*)} \frac{x^2}{x^2 + 1.5^2 
		\left( \frac{L}{L_*} \right)^{0.4}} 
	\right]^{1/2} {\rm km~s^{-1}}
\label{3-10}
\end{equation}
where $R_{opt} \equiv 3.2\,R_D$ is the optical radius and $R_D$ is the disc exponential length-scale; $x = r/R_{opt}$; $L$ is the luminosity. We would like to compare the rotation curve for the color fields \eqref{2-70} with the Universal Rotation Curve \eqref{3-30} where, for example, $L/L_* = 1$
\begin{equation}
	V_{URC} \left( \frac{r}{R_{opt}} \right) = 
	V(R_{opt}) \left[ 
		\frac{1.4184 \; x^{1.22}}{ \left( x^2 + 0.78^2 \right)^{1.43}} + 
		\frac{1.07251 \; x^2}{x^2 + 1.5^2} 
	\right]^{1/2} {\rm km~s^{-1}}.
\label{3-20}
\end{equation}
For the Dark Matter the Universal Rotation Curve is 
\begin{equation}
	V_{DM}^2 \left( \frac{r}{R_{opt}} \right) = 
	V^2(R_{opt}) \frac{1.07251 \; x^2}{x^2 + 1.5^2} \; 
	{\rm km~s^{-1}}.
\label{3-30}
\end{equation}
The profiles of $V_{URC}(x), V_{DM}^2(x), V_{LM}^2(x)$ in Fig.~\ref{fg3} are presented ($V_{DM}^2$ is the rotation curve for the Dark Matter, $V_{LM}^2(x)$ is the rotation curve for the light matter).
\par
At the center $r \approx 0$ the approximate solution has the form \eqref{2-20} and the mass density \eqref{2-60} approximately is 
\begin{equation}
	\rho(x) \approx \frac{6}{g^2 c^2 r_0^4} \frac{v_2^2}{r_0^4}.
\label{3-40}
\end{equation}
Consequently the rotation curve will be 
\begin{equation}
	V^2 \approx \frac{G \hbar}{c} \frac{1}{{g'}^2} \frac{v_2^2}{r_0^2}
	\left( \frac{r}{r_0} \right)^2 \; m~s^{-1} .
\label{3-50}
\end{equation}
The comparison with Eq.~\eqref{3-30} by $x \ll 1$ gives us 
\begin{equation}
	v_2 \approx 20 \sqrt{\frac{c}{G \hbar}} \; \; 
	g' \frac{V_{opt}}{R_{opt}} r_0^2 .
\label{3-60}
\end{equation}
If $r_0$ is very small in comparison with $R_{opt}$ then far away from the center the functions $v(x)$ and $w(x)$ are presented in Eq's \eqref{2-80}-\eqref{2-100} and the mass density is 
\begin{equation}
	\varepsilon_\infty(x) = 
	c^2 \rho_\infty(r) \approx \frac{2}{3} \frac{1}{g^2 r_0^4} 
	\alpha^2 \left( \alpha - 1 \right) \left( 3 \alpha - 1 \right) 
	\left( \frac{r}{r_0} \right)^{2 \alpha - 4}. 
\label{3-70}
\end{equation}
In this case we can estimate the values of square of speed in the following way 
\begin{eqnarray}
	V^2 &=& \frac{G \hbar}{c} \frac{1}{{g'}^2 r_0^2} \frac{1}{x} 
	\left( 
		\int \limits_0^{x_1} x^2 \varepsilon(x) dx + 
		\int \limits_{x_1}^x x^2 \varepsilon(x) dx
	\right) \approx 
	\left[
		\frac{G \hbar}{c} \frac{1}{{g'}^2 r_0^2} \frac{1}{x} 
		\int \limits_0^x x^2 \varepsilon_\infty(x) dx 
	\right] - V^2_0 ,
\label{3-73}\\
	V^2_0 &=& \frac{G \hbar}{c} \int \limits_0^{x_1} x^2 
	\left[ \varepsilon_\infty(x) - \varepsilon(x) \right] dx 
\label{3-76}
\end{eqnarray}
here for the region $x > x_1$ the asymptotical Eq.~\eqref{3-70} is valid. One can say that $V_0^2$ is a systematical error of equation 
\begin{equation}
	V^2 = \frac{G \hbar}{c} \frac{1}{{g'}^2 r_0^2} \frac{1}{x} 
	\int \limits_0^x x^2 \varepsilon_\infty(x) dx 
\label{3-78}
\end{equation}
and the numerical value of $V^2_0$ is defined near to the center of galaxy where according to Eq. \eqref{3-76} the difference $\varepsilon_\infty(x) - \varepsilon(x)$ is maximal. Thus the asymptotical behavior of the rotation curve for the domain filled with the SU(3) gauge field is 
\begin{equation}
	V^2 \approx \frac{2}{3} \frac{G \hbar}{c}  \frac{1}{{g'}^2 r_0^2} 
	\alpha^2 \left( \alpha - 1 \right) \left( 3 \alpha - 1 \right) 
	\left( \frac{r}{r_0} \right)^{2 \alpha - 2} 
	 - V^2_0 .
\label{3-80}
\end{equation}
In Fig.~\ref{fg4} we see that it is possible to choose the parameters $\alpha, r_0, g', V_0$ in such a way that we have very good coincidence of the Universal Rotation Curve \eqref{3-30} and the rotation curve for spherically symmetric distribution of SU(3) gauge field \eqref{1-30}-\eqref{1-90}. 

\section{Invisibility of color fields}

The invisibility of classical color fields is based on the fact that only \emph{colored} elementary particles may interact with the classical non-Abelian gauge fields. The equations describing the motion of a colored particle are Wong's equations 
\begin{eqnarray}
	m \frac{d^2 x^\mu}{ds^2} &=& -g F^{a \mu \nu} M^a \frac{d x^\nu}{ds} ,
\label{3-10}\\
	\frac{d M^a}{ds}  &=& -g f^{abc} \left(
		A^{b \mu} \frac{d x^\mu}{ds} 
	\right)M^c
\label{4-10}
\end{eqnarray}
where $x^\mu(s)$ is the 4D trajectory of the particle with the mass $m$, $M^a$ is the color  components of color charge of the particle, $\left( M^a \right)^2 = M^2 = const$. Now we see that the ordinary elementary particles do not interact with the color fields presented here as they are colorless in the consequence of confinement for strong interactions. 
\par
Only monopoles and dyons may interact with these fields. Another possibility to find the influence of the color fields on elementary particles is the following. Some particles (proton , neutron and so on) may have an inner structure, i.e. a color electric or magnetic dipole or quadrupole which will interact with the external inhomogeneous color field. But this interaction should be very small. 

\section{Conclusions and discussion}

We have shown that in principle the classical SU(3) color fields have weakly decreasing mass density that allows us to offer the corresponding gauge field distribution as a candidate for DM expaining the Universal Rotation Curve of spiral galaxies. The question about the adaptability of the presented model of DM for the explanation of galaxy movement in clusters and galaxy clusters in superclusters demands the separate consideration that will be done in our future investigations. The distinctive feature of this DM model is that it uses well established SU(3) Yang-Mills theory and there is not necessity neither to modify Newton's gravitational law or to introduce weakly interacting supersymmetrical particles. It is interesting that the invisibility of gauge DM is connected with the confinement in quantum chromodynamics. 
\par
Very important questions are: how big is the SU(3) domain, has it an infinite or finite 
volume ? In this connection it is necessary to do the next remark. The gauge fields presented here are strongly oscillating far away from the center (for $r_0 \ll R_{opt}$). On some distance from the center the oscillations become so strong that quantum effects begin essential: we should take into account the Heisenberg Uncertainty Principle. These effects appear when the quantum fluctuations of gauge field become comparable with the magnitude of the color fields on the distance about the period of oscillations. In this case we have to apply \emph{nonperturbative} quantization in order to describe the quantum color fields. Now we would like to note that in Ref.~\cite{Dzhunushaliev:2006di} the approximate model of a non-perturbative quantization is offered and the solution describing a ball filled with a gauge condensate is received. The quantum fields of the ball decrease asymptotically as $\frac{e^{-r/l_0}}{r}$ (here $l_0$ is a constant) and we will have a finite total mass of the SU(3) domain. Thus as a whole the picture of the color field distribution (which is DM in the presented model) looks as follows: there is a ball filled with the \emph{classical} color field which gives us the observable rotation curve and far enough from the center the gauge fields become \emph{quantum} one.

\end{document}